\documentclass[journal=jctcce,manuscript=article,layout=twocolumn]{achemso}

\usepackage[colorlinks,linkcolor=blue,citecolor=blue,urlcolor=black,bookmarks=false,hypertexnames=true]{hyperref} 
\usepackage[utf8]{inputenc}
\usepackage[version=3]{mhchem}
\usepackage{amssymb}
\usepackage{color}
\usepackage{braket}
\usepackage{xspace}
\usepackage{graphicx}
\usepackage{subfig}
\usepackage{threeparttable}
\usepackage{multirow}
\usepackage{textcomp}
\usepackage{caption}
\usepackage{array}
\usepackage{float}
\usepackage{cleveref}
\usepackage{xcolor}
\usepackage{geometry}
\newcolumntype{L}[1]{>{\raggedright\let\newline\\\arraybackslash\hspace{0pt}}m{#1}}
\newcolumntype{C}[1]{>{\centering\let\newline\\\arraybackslash\hspace{0pt}}m{#1}}
\newcolumntype{R}[1]{>{\raggedleft\let\newline\\\arraybackslash\hspace{0pt}}m{#1}}
\allowdisplaybreaks
\makeatletter
\let\l@addto@macro\relax
\makeatother
\usepackage[fontsize=11pt]{scrextend}

\let\oldmaketitle\maketitle
\let\maketitle\relax
\newcommand{\kc}{\textbf{k}}

\usepackage{afterpage}

\crefname{figure}{Figure}{Figures}
\crefname{subfigure}{Figure}{Figures}
\crefname{table}{Table}{Tables}
\crefname{equation}{Eq.}{Eqs.}
\crefname{section}{Section}{Sections}
\crefname{subsection}{Section}{Sections}

\author{Abdelrahman M.\@ Ahmed}
\affiliation{%
     Department of Chemistry and Biochemistry,	
     The Ohio State University,
     Columbus, Ohio 43210, United States
}
 \author{Alexander Yu.\ Sokolov}
 \email{sokolov.8@osu.edu}
 \affiliation{%
     Department of Chemistry and Biochemistry,
     The Ohio State University,
     Columbus, Ohio 43210, United States
 }

\begin{tocentry}
\includegraphics{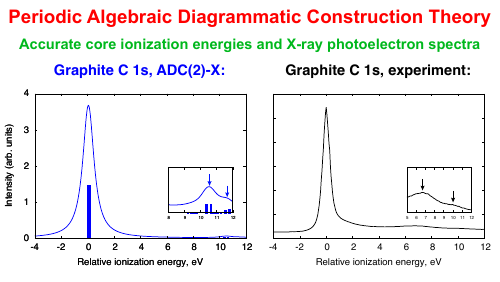}
\end{tocentry}

\title{{\color{blue}
    Core-Ionized States and X-ray Photoelectron Spectra of Solids From Periodic Algebraic Diagrammatic Construction Theory
}}

\begin{document}

%ArXiv %%%%%%%%%%%%%%
\newcommand*{\abstractext}{
We present the first-ever implementation and benchmark of periodic algebraic diagrammatic construction theory (ADC) for core-ionized states and X-ray photoelectron spectra (XPS) in crystalline materials.
Using a triple-zeta Gaussian basis set and accounting for finite-size and scalar relativistic effects, the strict and extended second-order ADC approximations (ADC(2) and ADC(2)-X) predict the core ionization energies of weakly correlated solids within $\sim$ 1.5 and 0.5 eV of experimental measurements, respectively.
We further demonstrate that the ADC(2)-X method can capture the satellite features in XPS spectra of graphite, cubic and hexagonal boron nitride, and \ce{TiO2}, albeit significantly overestimating their energies.
The ADC(2)-X calculations reveal that the satellite transitions display strong configuration interaction with excitations involving several frontier orbitals delocalized in phase space.
Our work demonstrates that ADC is a promising first-principles approach for simulating the core-excited states and X-ray spectra of materials, highlighting its potential and motivating further development. 
\vspace{0.25cm}
}
%ArXiv %%%%%%%%%%%%%%

%ArXiv %%%%%%%%%%%%%%
\twocolumn[
\begin{@twocolumnfalse}
\oldmaketitle
\vspace{-0.75cm}
\begin{abstract}
\abstractext
\end{abstract}
\end{@twocolumnfalse}
]
%ArXiv %%%%%%%%%%%%%%

\section{Introduction}

X-ray photoelectron spectroscopy (XPS) is one of the mainstay tools for investigating and characterizing crystalline materials.
By measuring the binding energies of electrons in localized core or inner-shell orbitals, XPS provides element-specific insights into the geometric and electronic structure at material surfaces\cite{sugaphotoelectron,Fadley.2010.10.1016/j.elspec.2010.01.006,Greczynski.2023.10.1038/s43586-023-00225-y,Gengenbach.2021.10.1116/6.0000682,Stevie.2020.10.1116/6.0000412} 
XPS is widely used to characterize the properties of semiconductors, battery and energy storage materials, and heterogeneous catalysts where it allows to determine the surface composition, element oxidation states, and the presence of defects or adsorbates.\cite{zhong2019mini,borgatti2016hard,kalha2021hard,weiland2016recent}
However, interpreting experimental XPS spectra is often challenging as the peak assignment can be complicated by overlapping features, shake-up effects, sample charging, surface contamination, and defects, making it difficult to extract precise chemical information.\cite{salmeron2018surfaces}
Overcoming these problems requires input from accurate theoretical calculations that can compute the XPS signatures of a material and correlate them with its electronic structure.

Meanwhile, accurately simulating the XPS spectra of materials has proven to be difficult, requiring sophisticated descriptions of electron correlation and orbital relaxation in high-energy, localized open-shell electronic states.\cite{Fadley.1973.10.1007/978-94-010-2630-7_2,Pham.2018.10.1021/acs.jpclett.7b01382}
%Although a wide range of accurate theoretical tools have been developed for simulating the XPS spectra of molecules, the range of accurate theoretical methodologies for simulating the solid-state XPS spectra is much more narrow. 
Widely used density functional theory (DFT) allows to efficiently capture weak electron correlation effects but suffers from the derivative discontinuity and self-interaction errors that become particularly significant in highly localized states.\cite{Olovsson.2010.10.1016/j.elspec.2009.10.007,Guille.2014.10.1063/1.4904720,Ozaki.2017.10.1103/physrevlett.118.026401,Vines.2018.10.1039/c7cp08503f,10.1103/physrevmaterials.3.100801,Kahk.2021.10.1021/acs.jpclett.1c02380}
To mitigate these problems, theoretical methods based on many-body perturbation theory have been developed.\cite{dmft1996,Olovsson.2009.10.1088/0953-8984/21/10/104205,Vinson:2011p115106,Giantomassi.2011.10.1002/pssb.201046094,Gulans.2014.10.1088/0953-8984/26/36/363202,Gilmore:2015p109,Vorwerk.2017.10.1103/physrevb.95.155121,Liang:2017p096402,Aoki.2018.10.1088/1361-648x/aabdfe,Golze:2018p4856,Setten.2018.10.1021/acs.jctc.7b01192,woicik2020charge,Golze.2020.10.1021/acs.jpclett.9b03423,Zhu.2021.10.1021/acs.jctc.0c00704,Roychoudhury.2023.10.1103/physrevb.107.035146}
In particular, the GW approximation\cite{Hedin:1965p796,Faleev:2004p126406,vanSchilfgaarde:2006p226402,Neaton:2006p216405,Samsonidze:2011p186404,vanSetten:2013p232,Reining:2017pe1344,Setten.2018.10.1021/acs.jctc.7b01192,Golze.2020.10.1021/acs.jpclett.9b03423,Zhu.2021.10.1021/acs.jctc.0c00704} provides significant improvements over DFT in capturing electron correlation effects and reducing the self-interaction errors while maintaining affordable computational cost.
Although the GW methods can compute accurate binding energies of valence electrons in weakly correlated materials, their calculations are usually performed using plane-wave basis sets that exhibit slow convergence for core-electron densities and are not well suited for simulating core-ionized states and XPS spectra.
Augmenting plane waves by other basis functions allows to ameliorate this problem but significantly complicates implementation.\cite{Ishii.2010.10.2320/matertrans.m2010303,Aoki.2018.10.1088/1361-648x/aabdfe}
For these reasons, most correlated calculations of periodic systems are performed by replacing the core electronic densities with pseudopotentials that prevent access to core excitation energies and wavefunctions. 

In contrast to plane waves, all-electron Gaussian basis sets can describe the properties of core and valence electrons on equal footing and have been widely used to compute accurate core-electron spectra of molecules with quantum chemical theories.\cite{Besley:2010p12024,Besley:2012p42,Wadey.2014.10.1021/ct500566k,Coriani:2015p181103,Tenorio.2016.10.1021/acs.jctc.6b00524,Norman:2018p7208,Tenorio:2019p224104,Liu:2019p1642,Vidal:2019p3117,Vidal:2020p8314,Besley.2020.10.1021/acs.accounts.0c00171,Besley.2021.10.1002/wcms.1527,Bintrim.2021.10.1063/5.0035141}
Recent developments enable Gaussian-based periodic simulations of crystalline materials using advanced electronic structure methods that systematically capture electron correlation effects, including some flavors of GW,\cite{Wilhelm.2017.10.1103/physrevb.95.235123,Golze:2018p4856,Golze.2020.10.1021/acs.jpclett.9b03423,Zhu.2021.10.1021/acs.jctc.0c00704,Lei.2022.10.1063/5.0125756,Pokhilko.2022.10.1063/5.0114080} coupled cluster theory,
\cite{Zhang:2019p123,Hirata:2004p2581,Katagiri:2005p224901,Booth:2013p365,McClain:2016p235139,McClain:2017p1209,Gao:2020p165138,Gallo:2021p064106,Wang:2020p3095,Wang.2021.10.1021/acs.jctc.1c00692,Furukawa.2018.10.1063/1.5029537} 
and embedding approaches.\cite{Cui.2019.10.1021/acs.jctc.9b00933,Zhu:2020p141,Rusakov:2019p229}
Among these methods, equation-of-motion coupled cluster theory (EOM-CC) has been shown to predict accurate ionization energies and bandgaps of weakly correlated materials.\cite{McClain:2016p235139,McClain:2017p1209,Gao:2020p165138,Gallo:2021p064106,Wang:2020p3095,Wang.2021.10.1021/acs.jctc.1c00692,Lange:2021p081101,Vo.2024.10.1063/5.0187856}
However, the capabilities of periodic EOM-CC methods remain significantly limited by their steep computational cost, posing challenges for large-scale simulations and complex materials.

In this work, we present an implementation and benchmark of periodic algebraic diagrammatic construction theory (ADC)\cite{Banerjee:2022p5337} for simulating core ionization energies and XPS spectra of crystalline systems. 
By leveraging the power of perturbation theory, ADC allows to accurately capture orbital relaxation and electron correlation effects with lower computational cost relative to the EOM-CC approximations.\cite{Schirmer:1982p2395,Schirmer:1983p1237,Angonoa:1987p6789,Mertins:1996p2140,Schirmer:1998p4734,Schirmer:2001p10621,Thiel:2003p2088,Trofimov:2005p144115,Sokolov:2018p204113,Dempwolff:2019p064108,Chatterjee:2020p6343,Dempwolff.2022.10.1063/5.0079047,Leitner.2024.10.1021/acs.jpca.4c03037,Banerjee:2019p224112,Banerjee:2021p074105,Banerjee:2023p3037}
Specifically, the strict and extended second-order ADC methods (ADC(2) and ADC(2)-X) combined with the core-valence separation (CVS) technique\cite{Cederbaum:1980p481,Cederbaum:1980p206,Angonoa:1987p6789,Thiel:2003p2088,Coriani:2015p181103,Herbst:2020p054114} are able to deliver accurate core ionization energies and XPS spectra for large molecules in a good agreement with experimental results.\cite{Wenzel:2015p214104,Ambroise.2021.10.1021/acs.jctc.1c00042,Moura:2022p4769,Moura:2022p8041,Gaba:2024p15927,deMoura:2024p5816}
Building on the recently developed periodic formulation of ADC for charged excitations,\cite{Banerjee:2022p5337} we extend the capabilities of ADC(2) and ADC(2)-X to simulate the core-ionized states and spectra in crystalline materials.

This paper is organized as follows.
First, we briefly review periodic ADC (\cref{sec:theory:overview_adc}), describe its combination with CVS for core-ionized states (\cref{sec:theory:periodic_adc_theory_cvs}) and implementation in the PySCF software package\cite{Sun:2020p024109} (\cref{sec:implementation}).
Following the presentation of computational details (\cref{sec:Computational Details}),  in \cref{sec:results:benchmark} we benchmark the accuracy of CVS-ADC(2) and CVS-ADC(2)-X for predicting the core ionization energies of weakly correlated materials against experimental data. 
Finally, in \cref{sec:results:satellites} we use CVS-ADC(2)-X to simulate the X-ray photoelectron spectra of graphite, cubic and hexagonal boron nitride, and rutile (\ce{TiO2}), paying particular attention to its ability in reproducing the satellite features.
Our conclusions and outlook of future developments are presented in \cref{sec:conclusions}.

\section{Methods}
\label{sec:Theory}

\subsection{Periodic ADC for Ionized Crystalline States}
\label{sec:theory:overview_adc}

We begin with a brief overview of periodic non-Dyson ADC for the ionized states and photoelectron spectra of solids that we developed in our previous work. 
For additional details, we refer the readers to Ref.\@ \citenum{Banerjee:2022p5337}.
Our starting point here is the one-particle Green’s function (1-GF) that describes the charged excitations (ionization, electron attachment) of a chemical system induced by a field with frequency $\omega$.\cite{Fetter:1971quantum,Dickhoff:2008many,Schirmer:2018} 
Limiting our discussion to the ionization processes, 1-GF can be expressed in reciprocal space as:
\begin{align}
G_{pq}(\omega,\kc) = \langle\Psi_{0}^{N}|a_{q\kc}^{\dagger} (\omega + H - E_{0}^{N})^{-1}a_{p\kc}|\Psi_{0}^{N}\rangle  \label{eq:G_pq}
\end{align}
where $\kc$ is crystal momentum, $H$ is the electronic Hamiltonian, $|\Psi_{0}^{N}\big\rangle$ and  $E_{0}^{N}$ are the exact ground-state wavefunction and energy of a crystalline  $N$-electron unit cell, respectively.
The single-particle states, represented by the fermionic creation ($a_{q\kc}^\dag$) and annihilation ($a_{p\kc}$) operators, are obtained from a periodic self-consistent field calculation (SCF).
Due to the translational symmetry of $H$ and crystal momentum conservation, 1-GF depends on only one  $\kc$ value at a time.

The 1-GF can be recast in the spectral (or Lehmann) representation\cite{Kobe:1962p448}
\begin{align}
    \textbf{G}(\omega,\kc) = \mathbf{\tilde{X}}(\kc)(\omega\textbf{1} - \boldsymbol{\tilde{\Omega}}(\kc))^{-1}\mathbf{\tilde{X}}^{\dagger}(\kc)  \label{eq:G_pq_matrix}
\end{align}
where each factor is related to a physical observable that can be measured in photoelectron spectroscopy.
In particular, $\boldsymbol{\tilde{\Omega}}(\kc)$ is the $\kc$-dependent diagonal matrix of vertical ionization energies $\tilde{\Omega}_n(\kc) = E_{0}^{N} - E_{n}^{N - 1}(\kc)$ while $\mathbf{\tilde{X}}(\kc)$ is the matrix of spectroscopic amplitudes $\tilde{X}_{qn}(\kc) = \langle\Psi_{0}^{N}|a_{q\kc}^\dag|\Psi_{n}^{N-1}(\kc)\rangle$ that describes ionization probabilities. 
Both $\boldsymbol{\tilde{\Omega}}(\kc)$ and $\mathbf{\tilde{X}}(\kc)$ depend on the $(N-1)$-electron eigenstates of ionized crystal $\ket{\Psi_{n}^{N-1}(\kc)}$ that are computationally expensive to calculate without introducing approximations.

To compute $\boldsymbol{\tilde{\Omega}}(\kc)$ and $\mathbf{\tilde{X}}(\kc)$ for realistic materials, in periodic non-Dyson ADC the 1-GF is written in a non-diagonal spectral form
\begin{align}
	\textbf{G}(\omega,\kc) = \textbf{T}(\kc)(\omega\textbf{1}-\textbf{M}(\kc))^{-1}\textbf{T}^{\dagger}(\kc)  \label{eq:G_approx}
\end{align}
which does not require {\it a priori} knowledge of exact eigenstates.
Similar to $\boldsymbol{\tilde{\Omega}}(\kc)$ and $\mathbf{\tilde{X}}(\kc)$ in \cref{eq:G_pq_matrix}, the effective Hamiltonian ($\textbf{M}(\kc)$) and transition moment ($\textbf{T}(\kc)$) matrices describe vertical ionization energies and transition probabilities but are expressed in a noneigenstate basis of orthogonal electronic configurations with $(N-1)$ electrons in a unit cell.

Separating the electronic Hamiltonian $H$ into the one-electron zeroth-order contribution (Fock operator $H^{(0)}$) and a  perturbation term (two-electron interaction $V$) allows to expand $\textbf{M}(\kc)$ and $\textbf{T}(\kc)$ in single-reference (M\o ller--Plesset) perturbation series
\begin{align} 
	\textbf{M$(\kc)$} &= \textbf{M$^{(0)}(\kc)$} +\textbf{M$^{(1)}(\kc)$} + ... + \textbf{M$^{(n)}(\kc)$} + ...\\
	\textbf{T$(\kc)$} &= \textbf{T$^{(0)}(\kc)$} +\textbf{T$^{(1)}(\kc)$} + ... + \textbf{T$^{(n)}(\kc)$} + ...
\end{align}
that are expected to converge rapidly as long as the reference SCF wavefunction is a good approximation to $|\Psi_{0}^{N}\big\rangle$  in \cref{eq:G_pq}.
Truncating $\textbf{M}(\kc)$ and $\textbf{T}(\kc)$ at perturbation order $n$ defines the $n$th-order ADC approximation (ADC($n$)).

Diagonalizing the effective Hamiltonian matrix $\textbf{M}(\kc)$ 
\begin{align}
	\textbf{M}(\kc)\textbf{Y}(\kc)=\textbf{Y}(\kc)\boldsymbol{\Omega}(\kc) \label{eq:eig}
\end{align}
yields the occupied band energies $\boldsymbol{\Omega}(\kc)$ and eigenvectors $\textbf{Y}(\kc)$ that can be used to calculate the spectroscopic amplitudes 
\begin{align}
	\textbf{X}(\kc)=\textbf{T}(\kc) \textbf{Y}(\kc)  \label{eq:spec_amp}
\end{align}
and transition probabilities (so-called spectroscopic factors)
\begin{align}
	\label{eq:spec_factors}
	P_{\mu}(\kc) = \sum_{p} |X_{ p\mu}(\kc)|^{2} 
\end{align}
Evaluating the ADC($n$) 1-GF in its spectral form
\begin{align}
	\label{eq:g_adc}
	\mathbf{G}(\omega,\kc) &= \mathbf{X}(\kc) \left(\omega \mathbf{1} - \boldsymbol{\Omega}(\kc)\right)^{-1} \mathbf{X}(\kc)^\dag 
\end{align}
provides access to momentum-resolved density of states
\begin{align}
	\label{eq:dos}
	A(\omega,\kc) &= -\frac{1}{\pi} \mathrm{Im} \left[ \mathrm{Tr} \, \mathbf{G}(\omega,\kc) \right] 
\end{align}

\subsection{Periodic ADC With Core-Valence Separation}
\label{sec:theory:periodic_adc_theory_cvs}

\begin{figure*}[t!]
	\centering
	\includegraphics[width=0.6\textwidth]{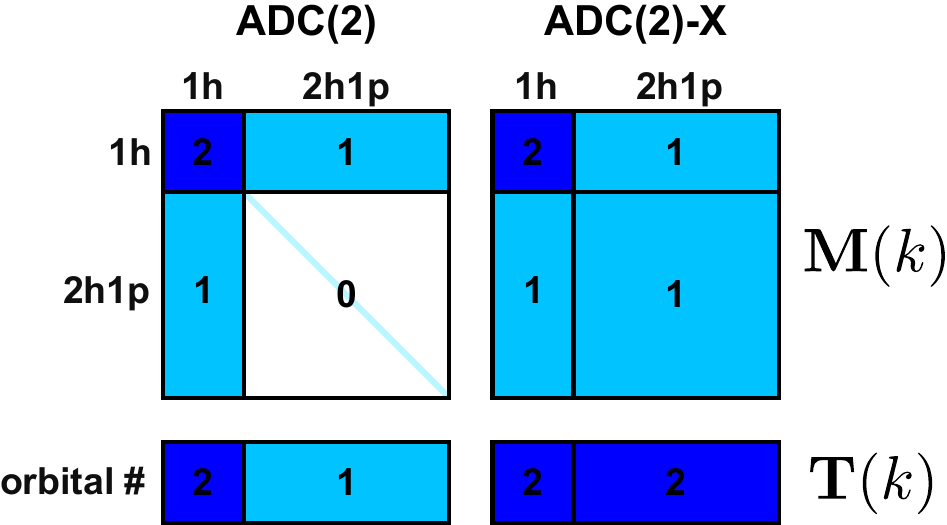}
	\caption{Perturbative structures of the effective Hamiltonian ($\mathbf{M}(k)$) and transition moments ($\mathbf{T}(k)$) matrices in the periodic ADC(2) and ADC(2)-X methods. 
		Numbers denote the perturbation order to which the effective Hamiltonian and transition moments are expanded for each sector. 
		Shaded areas indicate nonzero matrix elements.
		The one-hole (1h) and two-hole one-particle (2h1p) electronic configurations are depicted in \cref{fig:CVS_excitaton_manifold}.
	}
	\label{fig:ADC_matrices}
\end{figure*}

\begin{figure*}[t!]
    \centering
    \includegraphics[width=0.85\textwidth]{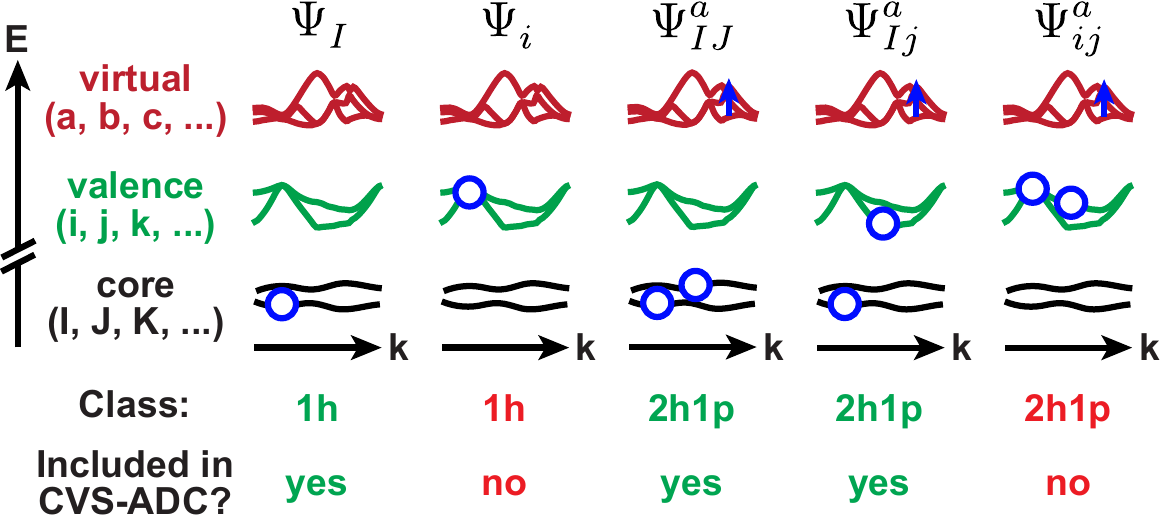}
    \caption{Illustration of one-hole (1h) and two-hole one-particle (2h1p) electronic configurations used to represent the effective Hamiltonian ($\mathbf{M}(k)$) and transition moments ($\mathbf{T}(k)$) matrices in the periodic ADC(2) and ADC(2)-X methods (\cref{fig:ADC_matrices}). 
    Holes and particles are depicted as circles and arrows, respectively.}
    \label{fig:CVS_excitaton_manifold}
\end{figure*}

To simulate the core-ionized states and XPS spectra of crystalline materials, we will employ two periodic ADC approximations: (i) the strict second-order ADC method (ADC(2)) and (ii) the extended second-order ADC variant (ADC(2)-X).
\cref{fig:ADC_matrices} shows the perturbative structure of effective Hamiltonian $\textbf{M}(\kc)$ and transition moments $\textbf{T}(\kc)$ matrices in these two methods.
Each matrix element is evaluated in the basis of $(N-1)$-electron configurations that belong to two classes: the zeroth-order one-hole (1h) states $\ket{\Phi_{\pm \mu}^{(0)}(\mathbf{k})}$ and the first-order two-hole one-particle configurations (2h1p) $\ket{\Phi_{\pm \mu}^{(1)}(\mathbf{k})}$ (\cref{fig:CVS_excitaton_manifold}).
The 1h excitations represent the primary ionization events that appear as intense peaks in photoelectron spectra while the 2h1p configurations describe the satellite (or shake-up) peaks with weaker intensities. 
As illustrated in \cref{fig:ADC_matrices}, ADC(2) and ADC(2)-X describe the 1h transitions up to the second order in M\o ller--Plesset perturbation theory (MP2)\cite{Moller:1934kp618} but differ in the treatment of 2h1p excitations that are incorporated to a higher order in the extended approximation. 

Out of a large number of 1h and 2h1p excitations described in ADC(2) and ADC(2)-X, only few correspond to the core or inner-shell ionization processes that are observed in the XPS spectra.
However, the high-energy core-ionized states are deeply embedded in the eigenstate spectrum of effective Hamiltonian matrix $\textbf{M}(\kc)$ and are expensive to compute by iteratively solving the ADC eigenvalue problem in \cref{eq:eig}.
Here, we employ the core--valence separation approximation (CVS)\cite{Cederbaum:1980p481,Cederbaum:1980p206} that allows to efficiently compute core-ionized states by neglecting their coupling with the ionized configurations in valence orbitals due to the large differences in their energies and spatial localization. 
The CVS approximation has been widely used to simulate the X-ray absorption and photoelectron spectra of molecules.\cite{Angonoa:1987p6789,Thiel:2003p2088,Coriani:2015p181103,Herbst:2020p054114,Tenorio.2016.10.1021/acs.jctc.6b00524,Norman:2018p7208,Tenorio:2019p224104,Liu:2019p1642,Vidal:2019p3117,Vidal:2020p8314,HelmichParis:2021pe26559,Peng:2019p1840,Moura:2022p4769,Gaba:2024p15927,deMoura:2024p5816}

To introduce CVS, the occupied periodic orbitals are separated into two groups: 1) ``core'' that includes all low-energy orbitals starting with the one that is expected to be ionized first in the XPS spectrum and 2) ``valence'' that incorporates the remaining (higher lying) occupied band orbitals (\cref{fig:CVS_excitaton_manifold}). 
Next, the ADC effective Hamiltonian matrix $\textbf{M}(\kc)$ is computed and diagonalized in the basis of 1h and 2h1p excitations involving at least one core orbital.
The resulting CVS-ADC(2) and CVS-ADC(2)-X methods have two important properties:
(i) they enable direct calculations of core-ionized states and XPS spectra using standard (iterative) eigenvalue solvers,
(ii) they are more computationally efficient than their non-CVS counterparts since the number of 1h and 2h1p configurations used to represent $\textbf{M}(\kc)$ and $\textbf{T}(\kc)$ is significantly reduced.
The CVS approximation has been shown to be very accurate for the K-edge ($1s^{-1}$) ionization of molecules,\cite{Herbst:2020p054114} although its accuracy for the higher lying core or inner-shell orbitals (e.g., L or M edges) has been difficult to assess.\cite{Peng:2019p1840,Banerjee:2019p224112,HelmichParis:2021pe26559}

It is important to point out that comparing the computed and experimental core ionization energies in gapped solids is complicated by the experimental uncertainty in their Fermi level, which can be anywhere in the band gap region and is highly sensitive to surface defects and impurities.\cite{Walter.2019.10.48550/arxiv.1902.02958}
To remove this uncertainty, it is common to adjust the core ionization energy (CIE) of a material by the first ionization potential corresponding to the valence band maximum (VBM):\cite{Aoki.2018.10.1088/1361-648x/aabdfe,Zhu.2021.10.1021/acs.jctc.0c00704}
\begin{equation}
	\label{eq:modified_cie}
	E_{\mathrm{core}} = E_{\mathrm{CIE}}  - E_\mathrm{VBM}
\end{equation}
The modified core ionization energy $E_{\mathrm{core}}$ allows to benchmark theoretical methods against experimental data avoiding the uncertainty in the Fermi level. 
We will use \cref{eq:modified_cie} to define the core ionization energy for our benchmark of CVS-ADC methods in \cref{sec:results:benchmark}.

\subsection{Periodic CVS-ADC Implementation}
\label{sec:implementation}

We implemented the $k$-adapted periodic CVS-ADC methods in the development version of PySCF software package.\cite{Sun:2020p024109} 
The CVS-ADC equations are solved in a one-electron crystalline molecular orbital basis
\begin{align}
	\psi_{p\kc}(\textbf{r}) = \sum_{\mu}c_{\mu p\kc} \phi_{\mu \kc}(\textbf{r})  \label{eq:crys_MO}
\end{align}
constructed as linear combinations of translation-symmetry-adapted Gaussian atomic orbitals 
\begin{align}
	\phi_{\mu\kc}(\textbf{r}) = \sum_{\textbf{T}}e^{i\kc\cdot\textbf{T}}\chi_{\mu}(\textbf{r} - \textbf{T})  \label{eq:crys_AO} 
\end{align}
where $\chi_{\mu}(\textbf{r} - \textbf{T})$ are the atom-centered Gaussian basis functions, $\textbf{T}$ is a lattice translation vector, and $\kc$ is a crystal momentum vector in the first Brillouin zone.\cite{McClain:2017p1209,Sun:2017p164119}
The expansion coefficients $c_{\mu p\kc}$ are obtained from a periodic Hartree--Fock calculation.  
The $k$-point CVS-ADC code was verified against two less efficient implementations: 1) the $\Gamma$-point CVS-ADC program that does not support translational symmetry but can be used to compute core-ionized states at the $\Gamma$-point of the Brillouin zone ($\textbf{k}=(0,0,0)$) using a real-valued Gaussian basis set and 2) the $k$-point ADC code where the CVS projector was used to access core excitations following the work by Coriani and Koch.\cite{Coriani:2015p181103}

Although introducing the CVS approximation lowers the computational cost of CVS-ADC(2) and CVS-ADC(2)-X relative to their non-CVS variants, calculating core-ionized states requires using large all-electron basis sets to accurately capture the polarization of core orbitals upon ionization. 
Expanding the basis set leads to a steep increase in disk and memory usage associated with storing the two-electron repulsion integrals ($v_{p\kc_{p}q\kc_{q}}^{r\kc_{r}s\kc_{s}}$) and the correlation amplitudes of effective Hamiltonian ($t_{i\kc_{i}j\kc_{j}}^{a\kc_{a}b\kc_{b}(1)}$).\cite{Banerjee:2022p5337}
To reduce the disk and memory storage requirements of our CVS-ADC code, we implemented an integral-direct algorithm to efficiently calculate and handle $v_{p\kc_{p}q\kc_{q}}^{r\kc_{r}s\kc_{s}}$ and $t_{i\kc_{i}j\kc_{j}}^{a\kc_{a}b\kc_{b}(1)}$ using Gaussian density fitting, discarding them immediately after use.\cite{Sun:2017p164119,Bintrim.2022.10.1021/acs.jctc.2c00640} 
Additional savings were achieved by storing the two-index density-fitted integrals in a single-precision format as described in Ref. \citenum{Pokhilko:2018p4088}.
To ensure numerical stability, the matrix-vector products $\boldsymbol{\sigma}(\kc) = \bf{M(\kc)Y(\kc)}$ computed while iteratively solving the CVS-ADC eigenvalue problem (\cref{eq:eig}) were stored with double precision.

\subsection{Computational Details}
\label{sec:Computational Details}

\begin{table*}[t!]
	\captionsetup{justification=raggedright,singlelinecheck=false,font=footnotesize}
	\caption{
		Parameters of periodic ADC calculations performed for the benchmark study in \cref{sec:results:benchmark}. 
		The basis sets used for calculating core ionization energies and valence band maxima are denoted as basis set (CVS) and basis set (VBM), respectively. 
	}
	\label{tab:solids_benchmark}
	\setlength{\extrarowheight}{2pt}
	\setstretch{1}
	\scriptsize
	\centering
	\hspace*{-0.8cm}
	\begin{threeparttable}
		\begin{tabular}{lllll}
			\hline\hline
			Material  & Core & Basis set (CVS) & Basis set (VBM) & Extrapolated $k$-meshes \\\hline
			MgO	   	   & Mg 2p & cc-pwCVTZ	& cc-pwCVTZ	&	3$\times$3$\times$3, 4$\times$4$\times$4 \\
			MgO	   	   & O 1s & cc-pwCVTZ	& cc-pwCVTZ	&	3$\times$3$\times$3, 4$\times$4$\times$4 \\
			AlN	         & Al 2p      & cc-pwCVTZ(Al), cc-pVDZ(N) &  cc-pVTZ	& 3$\times$3$\times$2, 4$\times$4$\times$2 \\
			AlN          & N 1s       & cc-pwCVTZ(N), cc-pVDZ(Al) &  cc-pVTZ	& 3$\times$3$\times$2, 4$\times$4$\times$2 \\
			AlP	          & Al 2p  &cc-pwCVTZ	& cc-pwCVTZ&	3$\times$3$\times$3, 4$\times$4$\times$4\\
			Si	           &  Si  2p &cc-pwCVTZ	& cc-pwCVTZ&	3$\times$3$\times$3, 4$\times$4$\times$4\\
			BeO	        &  Be 1s      &cc-pwCVTZ(Be), cc-pVDZ(O) 	&cc-pVTZ &	3$\times$3$\times$2, 4$\times$4$\times$2\\
			BeO	         & O 1s      &cc-pwCVTZ(O), cc-pVDZ(Be) 	&cc-pVTZ &	3$\times$3$\times$2, 4$\times$4$\times$2\\
			cBN	         & B 1s   &cc-pwCVTZ	& cc-pwCVTZ&	3$\times$3$\times$3, 4$\times$4$\times$4\\
			cBN	         & N 1s   &cc-pwCVTZ	& cc-pwCVTZ&	3$\times$3$\times$3, 4$\times$4$\times$4\\
			SiC	          &  C 1s  &cc-pwCVTZ	& cc-pwCVTZ&	3$\times$3$\times$3, 4$\times$4$\times$4\\
			diamond	 &  C 1s  &cc-pwCVTZ	& cc-pwCVTZ&	3$\times$3$\times$3, 4$\times$4$\times$4\\
			GaN	        &  N 1s &cc-pwCVTZ(N), cc-pVDZ(Ga) &	cc-pVTZ &	3$\times$3$\times$2, 4$\times$4$\times$2\\
			ZnO	         & O 1s  &cc-pwCVTZ(O), cc-pVDZ(Zn)	&ADC(2): cc-pVTZ 	&	3$\times$3$\times$2, 4$\times$4$\times$2\\
				 &      &	&ADC(2)-X: cc-pwCVTZ(O), cc-pVDZ(Zn) 	&	\\\hline\hline
		\end{tabular}
	\end{threeparttable}
\end{table*}

To assess the accuracy of periodic CVS-ADC(2) and CVS-ADC(2)-X methods in simulating the core-ionized states and X-ray photoelectron spectra of solids, we performed two sets of benchmark calculations.
First, in \cref{sec:results:benchmark} we compute the core ionization energies of ten crystalline materials (MgO, AlN, AlP, Si, BeO, cubic BN = cBN, SiC, diamond, GaN, and ZnO) and compare them to available experimental data. 
Here, we adjust the core ionization energies (CIE) by the corresponding valence band maxima (VBM, \cref{eq:modified_cie}), which were computed using the non-CVS ADC methods.
\cref{tab:solids_benchmark} reports the basis sets and $k$-point grids employed in these calculations.
The structural parameters and lattice constants are reported in the Supplementary Information. 

For crystals with cubic lattices (MgO, AlP, Si, cBN, SiC, diamond), all ADC calculations were performed using the all-electron cc-pwCVTZ basis set\cite{Dunning:1989p1007,Woon:1995p4572,Peterson.2002.10.1063/1.1520138,Balabanov:2005p064107} with the 3$\times$3$\times$3 and 4$\times$4$\times$4 Monkhorst--Pack $k$-point grids. 
For materials with hexagonal lattices (AlN, BeO, GaN, ZnO) the CIE calculations were carried out using the cc-pwCVTZ basis set for the element being ionized and the cc-pVDZ basis for the other element.
The accuracy of such mixed basis set was benchmarked for cubic crystals (see Supplementary Information) showing errors of less than 0.1 eV in CIE relative to the calculations with the cc-pwCVTZ basis set. 
To compute the VBM energies of AlN, BeO, and GaN, we employed the all-electron cc-pVTZ basis set.
For ZnO, the ADC(2) VBM energy was computed using the cc-pVTZ basis while the mixed (cc-pwCVTZ(O), cc-pVDZ(Zn)) basis set was used to calculate the ADC(2)-X VBM to reduce computational cost.
The periodic ADC calculations for AlN, BeO, GaN, and ZnO were performed using the 3$\times$3$\times$2 and 4$\times$4$\times$2 $k$-point meshes.
 
To compare to the experimental data, CIE and VBM were extrapolated to the thermodynamic limit assuming that the finite-size error decays as $\mathcal{O}(N_k)^{-1/3}$.
Such extrapolation scheme was found to be accurate in our previous work.\cite{Banerjee:2022p5337}
The extrapolated core ionization energies adjusted by VBM (\cref{eq:modified_cie}) were supplied with the scalar relativistic corrections that were computed using CVS-ADC(2) with the X2C-1e relativistic Hamiltonian\cite{Dyall.2001.10.1063/1.1413512,Kutzelnigg:2005p241102,Liu:2009p031104} and 1$\times$1$\times$1 $k$-point mesh (see Supplementary Information). 
We have verified that the scalar relativistic corrections ($\Delta$X2C) exhibit negligible dependence on the $k$-point mesh and the level of theory.
The final core ionization energies computed as extrapolated $E_{\mathrm{CIE}} - E_{\mathrm{VBM}} + E_{\mathrm{\Delta X2C}}$ were benchmarked against the experimental data.

\begin{table*}[t!]
	\captionsetup{justification=raggedright,singlelinecheck=false,font=footnotesize}
	\caption{
		Parameters of periodic CVS-ADC calculations performed for the study of satellites in XPS spectra in \cref{sec:results:satellites}. 
	}
	\label{tab:solids_satellite}
	\setlength{\extrarowheight}{2pt}
	\setstretch{1}
	\scriptsize
	\centering
	\hspace*{-0.8cm}
	\begin{threeparttable}
		\begin{tabular}{lllll}
			\hline\hline
			Material  	  & Core& Basis set & $k$-mesh & Number of states \\\hline
			graphite  	  & C 1s & cc-pwCVDZ  & 3$\times$3$\times$2 & 70 \\
			cBN 			& B 1s & cc-pwCVDZ	& 3$\times$3$\times$3 & 140 \\
			hBN 			& B 1s & cc-pwCVDZ	& 3$\times$3$\times$2 & 100\\
			\ce{TiO2} 	 & Ti 2s & cc-pVDZ		& 2$\times$2$\times$1 & 140 \\
			\hline\hline
		\end{tabular}
	\end{threeparttable}
\end{table*}

In \cref{sec:results:satellites}, we investigate the accuracy of CVS-ADC(2)-X in simulating the satellite peaks in X-ray photoelectron spectra (XPS) of crystalline graphite, cubic and hexagonal boron nitrides (cBN and hBN), and rutile (\ce{TiO2}).
The parameters of these calculations are reported in \cref{tab:solids_satellite}.
Here, we focus on the relative energies and intensities between the intense edge peak and the weak satellite features.
Consequently, the computed core ionization energies were not extrapolated to the thermodynamic limit and did not incorporate the VBM and $\Delta$X2C corrections.
The XPS spectra for graphite, cBN, and hBN were simulated using the cc-pwCVDZ basis set while the \ce{TiO2} calculations employed the cc-pVDZ basis. 
The corresponding $k$-meshes and the number of states included in each calculation are provided in \cref{tab:solids_satellite}. 
The broadening parameters were selected to reproduce the width of the first peak in the experimental XPS spectra.

Since the computed core ionization energies and XPS spectra show very weak $k$-dependence, we only present the results calculated at the $\Gamma$-point ($\mathbf{k} = (0,0,0)$).
All periodic calculations were performed using Gaussian density fitting.\cite{Whitten:1973p4496,Dunlap:1979p3396,Vahtras:1993p514,Feyereisen:1993p359,Sun:2017p164119}
For all materials except MgO and BeO, the basis sets specified in \cref{tab:solids_benchmark,tab:solids_satellite} were supplied with the corresponding JKFIT and RIFIT auxiliary basis sets in the periodic SCF and ADC calculations, respectively.\cite{bross2013a,hattig2005a,weigend2002a,hill2008b}
In the case of MgO and BeO, the auxiliary basis sets were autogenerated using PySCF.
The experimental spectra were reproduced using WebPlotDigitizer.\cite{WebPlotDigitizer}

For brevity, we refer to CVS-ADC as ADC henceforth.

\section{Results \& Discussion}
\label{sec:results}

\subsection{Benchmark of Core Ionization Energies}
\label{sec:results:benchmark}

\begin{table*}[t!]
	\captionsetup{justification=raggedright,singlelinecheck=false,font=footnotesize}
	\caption{
		Core-level ionization energies (eV) computed using ADC(2) and measured experimentally.
		Results are shown for two $k$-point meshes, extrapolated thermodynamic limit (TDL), and the TDL values with valence band maximum and scalar relativistic corrections (TDL$-$VBM+$\Delta$X2C).
		See \cref{sec:Computational Details} and Supporting Information for additional details, including the choice of basis sets, and the VBM and $\Delta$X2C values.
		Also shown are mean absolute errors (MAE), maximum absolute errors (MAX), and standard deviations (STD) computed with respect to the experimental data. 
	}
	\label{tab:adc2_benchmark}
	\setlength{\extrarowheight}{2pt}
	\setstretch{1}
	\scriptsize
	\centering
	\hspace*{-0.8cm}
	\begin{threeparttable}
		\begin{tabular}{lcccccc}
			\hline\hline
			Core    & Material  & ADC(2) & ADC(2) & ADC(2) & ADC(2)  &      Experiment       \\
			&           &   3$\times$3$\times$2/3$\times$3$\times$3    &  4$\times$4$\times$2/4$\times$4$\times$4   &  TDL &  TDL$-$VBM+$\Delta$X2C  &               \\ \hline
			Mg 2p	& MgO	    &  35.86 	& 36.82	   &  39.69	   & 46.90	    & 46.71,\cite{Liu.2016.10.1016/j.vacuum.2016.10.012} 46.79\cite{Craft.2007.10.1063/1.2785022} \\
			Al 2p	& AlN	    &  59.44 	& 59.94	   &  62.30	    & 71.54	    & 70.56\cite{Veal.2008.10.1063/1.3032911} \\
			Al 2p	& AlP	    &  62.73 	& 63.33	   &  65.12	   		 & 73.17	    & 72.43\cite{Waldrop.1993.10.1116/1.586491} \\
			Si 2p	& Si	    &  87.90	    & 88.50	   &  90.31	   			& 100.10	& 98.95\cite{Yu.1990.10.1063/1.102747} \\
			Be 1s	& BeO	    &  101.95	& 102.58   &  105.52   & 111.20	& 109.8\cite{Hamrin.1970.10.1088/0031-8949/1/5-6/018} \\
			B 1s	& cBN	    &  174.28	& 175.26   &  178.20   		& 191.13	& 188.4\cite{Hamrin.1970.10.1088/0031-8949/1/5-6/018} \\
			C 1s	& SiC	    &  266.29	& 267.33   &  270.47  		& 283.56	& 281.45\cite{Waldrop.1990.10.1063/1.102744} \\
			C 1s	& diamond	&  266.05	& 267.14   &  270.43   & 286.33	& 283.7, 283.9\cite{chiang1989electronic} \\
			N 1s	& AlN	    &  381.86	& 382.66   &  386.48   & 395.94	& 393.87\cite{Veal.2008.10.1063/1.3032911} \\
			N 1s	& GaN	    &  378.52	& 379.77   &  385.72   & 397.58	& 395.2\cite{Duan.2013.10.1063/1.4807736} \\
			N 1s	& cBN	    &  379.42	& 380.80   &  384.92   		& 398.00	& 396.1\cite{Hamrin.1970.10.1088/0031-8949/1/5-6/018} \\
			O 1s	& MgO	    &  514.16	& 515.49   &  519.47   		& 527.03	& 527.28\cite{Chellappan.2013.10.1016/j.apsusc.2013.08.021} \\
			O 1s	& ZnO	    &  512.60	& 513.31   &  516.66  	 & 529.02	& 527.45\cite{Veal.2008.10.1063/1.3032911} \\
			O 1s	& BeO	    &  516.49	& 517.28   &  521.03 	  & 527.03	& 527.7\cite{Hamrin.1970.10.1088/0031-8949/1/5-6/018} \\\hline
			MAE             &                                      &   &  &  &               1.47 &  \\
			STD             &                                      &   &  &  &               1.05 &  \\
			MAX             &                                      &   &  &  &               2.73 &  \\ \hline\hline
		\end{tabular}
	\end{threeparttable}
\end{table*}

\begin{table*}[t!]
	\captionsetup{justification=raggedright,singlelinecheck=false,font=footnotesize}
	\caption{
		Core-level ionization energies (eV) computed using ADC(2)-X and measured experimentally.
		Results are shown for two $k$-point meshes, extrapolated thermodynamic limit (TDL), and the TDL values with valence band maximum and scalar relativistic corrections (TDL$-$VBM+$\Delta$X2C).
		See \cref{sec:Computational Details} and Supporting Information for additional details, including the choice of basis sets, and the VBM and $\Delta$X2C values.
		Also shown are mean absolute errors (MAE), maximum absolute errors (MAX), and standard deviations (STD) computed with respect to the experimental data. 
	}
	\label{tab:adc2x_benchmark}
	\setlength{\extrarowheight}{2pt}
	\setstretch{1}
	\scriptsize
	\centering
	\hspace*{-0.8cm}
	\begin{threeparttable}
		\begin{tabular}{lcccccc}
			\hline\hline
			Core    & Material  & ADC(2)-X & ADC(2)-X & ADC(2)-X & ADC(2)-X  &      Experiment      \\
			&           &   3$\times$3$\times$2/3$\times$3$\times$3    &  4$\times$4$\times$2/4$\times$4$\times$4   &  TDL &  TDL$-$VBM+$\Delta$X2C  &               \\ \hline
			Mg 2p	& MgO	& 36.18	& 37.06	& 39.71	& 46.34	& 46.71,\cite{Liu.2016.10.1016/j.vacuum.2016.10.012} 46.79\cite{Craft.2007.10.1063/1.2785022} \\
			Al 2p	& AlN	    & 59.78	& 60.25	& 62.47	& 70.81	& 70.56\cite{Veal.2008.10.1063/1.3032911} \\
			Al 2p	& AlP	    & 63.18	& 63.79	& 65.62	& 72.34	& 72.43\cite{Waldrop.1993.10.1116/1.586491} \\
			Si 2p	& Si	     & 88.41	& 89.07	& 91.06	& 99.01	& 98.95\cite{Yu.1990.10.1063/1.102747} \\
			Be 1s	& BeO	 & 101.18	& 101.77	& 104.53	& 109.59	& 109.8\cite{Hamrin.1970.10.1088/0031-8949/1/5-6/018} \\
			B 1s	& cBN	   & 173.46	& 174.43	& 177.36	& 188.87	& 188.4\cite{Hamrin.1970.10.1088/0031-8949/1/5-6/018} \\
			C 1s	& SiC	    & 266.12	& 267.08	& 269.96	& 281.33	& 281.45\cite{Waldrop.1990.10.1063/1.102744} \\
			C 1s	& diamond	& 265.74	& 266.79	& 269.95	& 283.91	& 283.7, 283.9\cite{chiang1989electronic} \\
			N 1s	& AlN	    & 382.41	& 383.06	& 386.15	& 394.70	& 393.87\cite{Veal.2008.10.1063/1.3032911} \\ 
			N 1s	& GaN	    & 379.46	& 380.45	& 385.10	& 396.21	& 395.2\cite{Duan.2013.10.1063/1.4807736} \\
			N 1s	& cBN	    & 379.95	& 381.13	& 384.69	& 396.36	& 396.1\cite{Hamrin.1970.10.1088/0031-8949/1/5-6/018} \\
			O 1s	& MgO	   & 515.82	& 516.84	& 519.91	& 526.89	& 527.28\cite{Chellappan.2013.10.1016/j.apsusc.2013.08.021} \\
			O 1s	& ZnO	    & 514.31	& 514.83	& 517.26	& 528.30	& 527.45\cite{Veal.2008.10.1063/1.3032911} \\
			O 1s	& BeO	    & 517.90	& 518.49	& 521.29	& 526.66	& 527.7\cite{Hamrin.1970.10.1088/0031-8949/1/5-6/018} \\\hline
			MAE     &  &    &  &  &    0.44       &  \\
			STD     &   &   &  &  &     0.56      & \\ 
			MAX     &  &   &  &  &    1.04      &  \\ \hline\hline
		\end{tabular}
	\end{threeparttable}
\end{table*}

\begin{figure*}[t!]
	\includegraphics[width=0.7\textwidth]{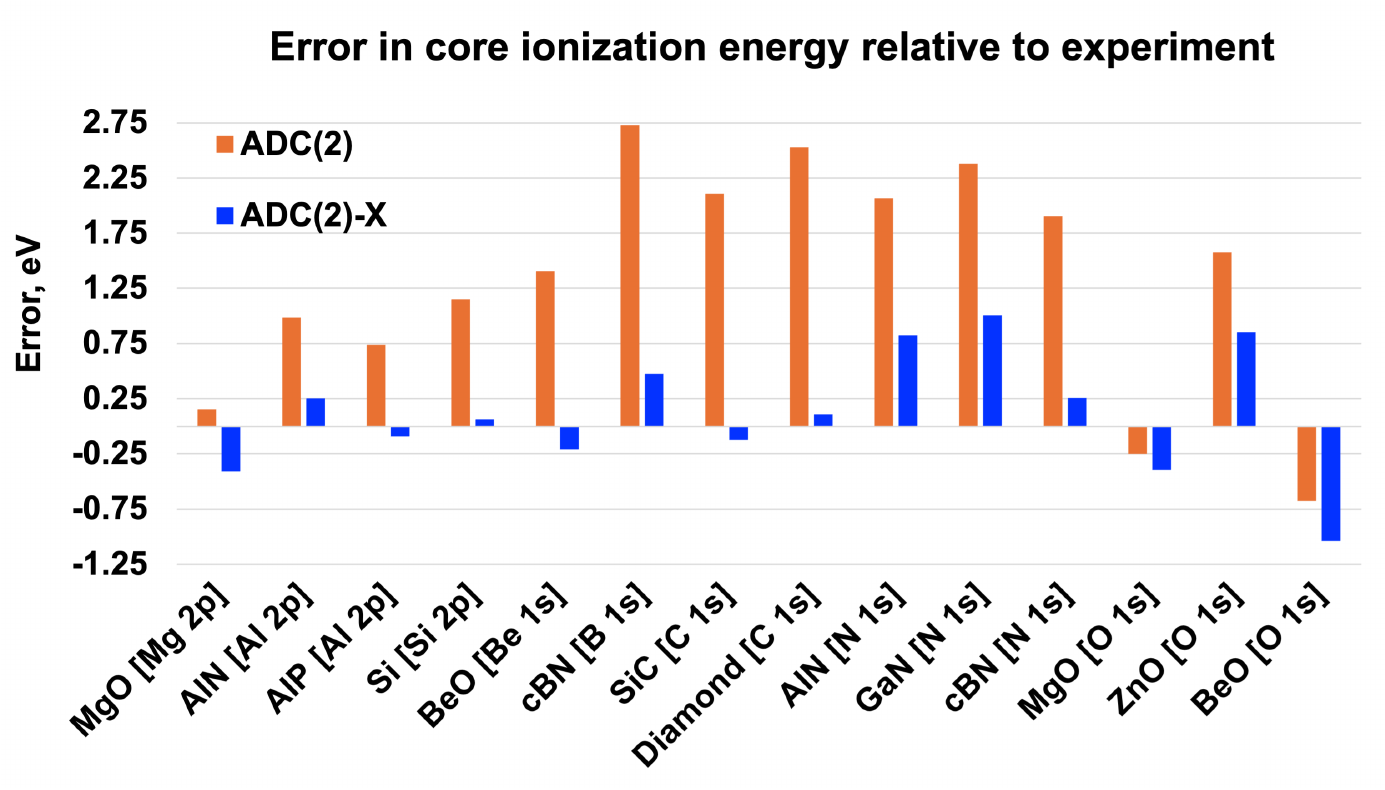}
	\caption{
		Errors in core-level ionization energies (eV) computed using ADC(2) and ADC(2)-X relative to experimental data.
		The core ionization energies were extrapolated to thermodynamic limit and include the valence band maximum and scalar relativistic corrections (TDL$-$VBM+$\Delta$X2C), see \cref{tab:adc2_benchmark,tab:adc2x_benchmark} for additional details.
	}.
	\label{fig:benchmark_barchart}
\end{figure*}

\cref{tab:adc2_benchmark,tab:adc2x_benchmark} report the core ionization energies of MgO, AlN, AlP, Si, BeO, cubic BN (cBN), SiC, diamond, GaN, and ZnO computed using ADC(2) and ADC(2)-X.
The results are presented for two $k$-meshes (3$\times$3$\times$2/3$\times$3$\times$3, 4$\times$4$\times$2/4$\times$4$\times$4), extrapolated thermodynamic limit (TDL), and the TDL values adjusted by the valence band maxima (VBM) and scalar relativistic corrections ($\Delta$X2C). 
The errors in TDL$-$VBM+$\Delta$X2C energies relative to experimental data are shown in \cref{fig:benchmark_barchart}.

The best agreement with experimental data is shown by the ADC(2)-X method, which predicts core ionization energies with the mean absolute error (MAE) of 0.44 eV, the standard deviation (STD) of 0.56 eV, and the maximum absolute error of 1.04 eV (MAX).
As shown in \cref{fig:benchmark_barchart}, the ADC(2)-X errors tend to increase with increasing magnitude of core ionization energy, reaching its maximum values ($\sim$ 1 eV) for the N 1s ionization in AlN and GaN, as well as the O 1s ionization in ZnO and BeO.
The ADC(2) method tends to overestimate core ionization energies with significantly larger MAE (1.47 eV), STD (1.05 eV), and MAX (2.73 eV).
In contrast to ADC(2)-X, the ADC(2) errors reach large values ($>$ 1.75 eV) for the B 1s core-ionized states of cBN, the C 1s states of SiC and diamond, and the N 1s ionization of AlN, GaN, and cBN, but are smaller for O 1s ionized states. 
Importantly, both ADC(2) and ADC(2)-X correctly predict the order of core-ionized states in increasing binding energy, with the exception of O 1s ionization in MgO, ZnO, and BeO.
In this case, the experimentally observed trend $E$(O 1s MgO) $<$ $E$(O 1s ZnO) $<$ $E$(O 1s BeO) is violated due to the ADC methods significantly overestimating $E$(O 1s ZnO) and underestimating $E$(O 1s BeO) (\cref{fig:benchmark_barchart}).
The challenges with reproducing the experimentally measured relative energies of these O 1s states have been also reported by Zhu and Chan in their Gaussian-based G$_0$W$_0$/DFT calculations using the cc-pCVTZ basis set.\cite{Zhu.2021.10.1021/acs.jctc.0c00704}

Overall, our results demonstrate that the ADC(2)-X method is significantly more accurate than ADC(2) for calculating the core ionization energies of crystalline materials.
This finding is consistent with the benchmark results for molecular core ionization energies where ADC(2)-X was found to outperform ADC(2).\cite{Trofimov:2005p144115,Moura:2022p8041,Moura:2022p4769}
For example, in a recent benchmark study of 16 polyatomic molecules, the MAE of ADC(2) and ADC(2)-X in core ionization energies were reported to be 1.26 and 0.84 eV, respectively, when using the cc-pCVTZ-X2C basis set.\cite{Moura:2022p8041}
The larger difference in the MAE of ADC(2) and ADC(2)-X observed in our solid-state benchmark (MAE = 1.47 and 0.44 eV, respectively) may, at least in part, originate from the higher accuracy of ADC(2)-X in predicting the crystalline VBM energies and fundamental band gaps compared to ADC(2), as demonstrated in Ref.\@ \citenum{Banerjee:2022p5337}.
Finally, we note that the ADC(2)-X performance compares favorably to the accuracy of Gaussian-based G$_0$W$_0$/DFT calculations performed by Zhu and Chan that reported MAE ranging from 0.57 to 7.92 eV depending on the density functional and the amount of Hartree--Fock exchange used.\cite{Zhu.2021.10.1021/acs.jctc.0c00704}
This suggests that the ADC(2)-X method can be used as a benchmark for G$_0$W$_0$/DFT and other lower-level theories when accurate experimental results are not available. 

\subsection{Satellite Peaks in X-ray Photoelectron Spectra}
\label{sec:results:satellites}

As discussed in \cref{sec:theory:overview_adc}, the ADC methods provide access to momentum-resolved density of states (\cref{eq:dos}) that can be used to simulate solid-state X-ray photoelectron spectra (XPS).
In addition to one-electron (or one-hole, 1h) ionization events that appear as intense (primary or edge) peaks, the XPS spectra exhibit satellite (shake-up or two-hole one-particle, 2h1p) transitions with weaker intensities occurring at higher energies.
Satellite excitations provide crucial insight into many-body interactions and electronic correlation effects in materials but are often not well understood due to the shortcomings of commonly employed theoretical methods (e.g., density functional theory, G$_0$W$_0$) that heavily rely on single-particle picture of ionization.
Both ADC(2) and ADC(2)-X incorporate the 2h1p excitations (\cref{fig:CVS_excitaton_manifold}) treating them to zeroth and first order in the effective Hamiltonian matrix, respectively (\cref{fig:ADC_matrices}).
Here, we employ ADC(2)-X to simulate the satellite transitions in X-ray photoelectron spectra of four representative materials---graphite, cubic boron nitride (cBN), hexagonal boron nitride (hBN), and rutile (\ce{TiO2})---and benchmark its accuracy against available experimental data.
The details of XPS calculations are described in \cref{sec:Computational Details}.

\begin{figure*}[t!]
	\centering
	\captionsetup{justification=raggedright,singlelinecheck=false,font=footnotesize}
	\includegraphics[width=0.7\textwidth]{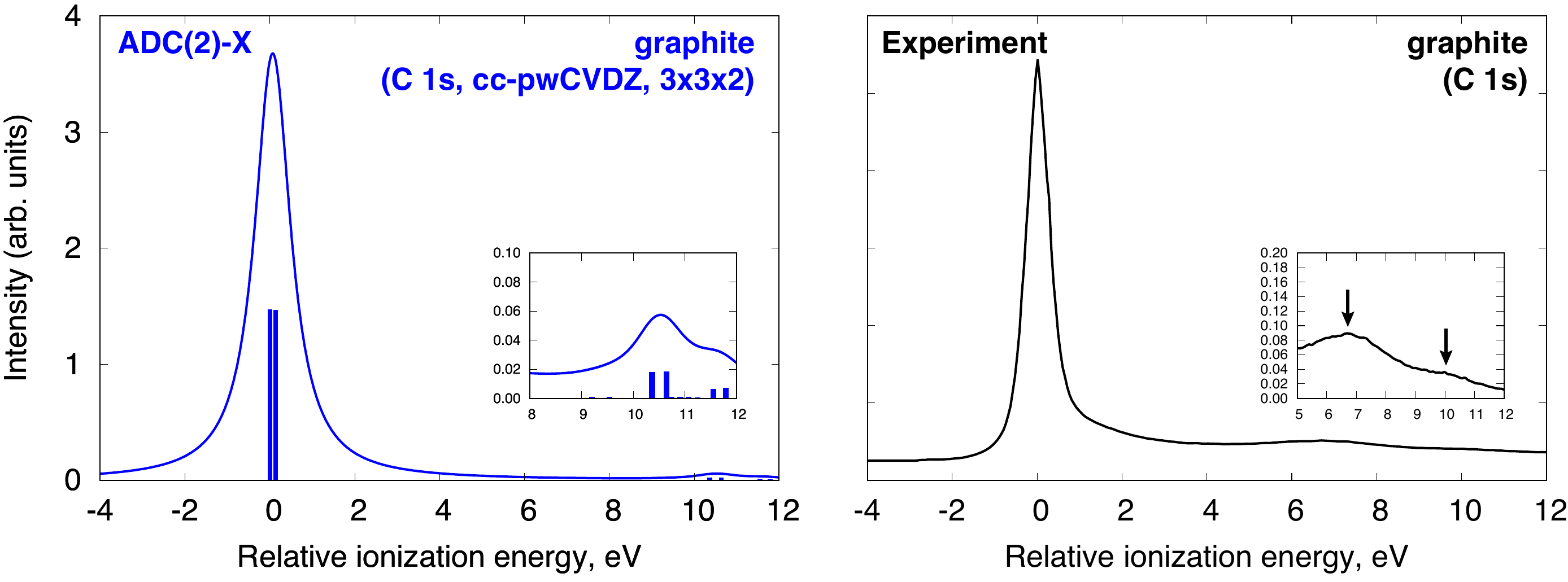}
	\caption{
		C 1s X-ray photoelectron spectrum of graphite computed using ADC(2)-X with the 3$\times$3$\times$2 $k$-mesh and the cc-pwCVDZ basis set.
		Also shown is experimental spectrum from Ref.\@ \citenum{Morgan.2021.10.3390/c7030051} with satellite features indicated by arrows.
		To align the position of first peak, the simulated and experimental spectra were shifted by $-$277.52 and $-$284.48 eV, respectively.
	}
	\label{fig:graphite_xps}
\end{figure*}

\cref{fig:graphite_xps} compares the simulated and experimental\cite{Morgan.2021.10.3390/c7030051} XPS spectra of graphite at the C 1s edge.
The experimental spectrum shows two broad satellite features: a peak at $\sim$ 6.6 eV and a shoulder at $\sim$ 10 eV relative to the primary C 1s edge peak.
When computed using the cc-pwCVDZ basis set and 3$\times$3$\times$2 $k$-mesh, the ADC(2)-X spectrum exhibits two shake-up features at $\sim$ 10.5 and 11.5 eV, in a qualitative agreement with experimental data.
Although the energies of these satellites are significantly overestimated, their relative intensities ($\sim$ 2:1 ratio) agree well with the experimental XPS spectrum. 
The ADC(2)-X calculations reveal that the satellite features consist of several transitions with non-zero photoelectron probabilities: four excitations with relatively high intensity (10.36, 10.64, 11.55, and 11.79 eV) and at least eight transitions that are $\sim$ 20 to 30 times less intense  (in the 9.20 to 11.24 eV range).
Analyzing the eigenvectors of ADC(2)-X effective Hamiltonian provides information about the orbital nature of each shake-up.
In all satellite transitions the C 1s ionization is accompanied by $\pi\rightarrow\pi^*$ excitations involving the two highest occupied and two lowest unoccupied band orbitals in a unit cell.
Each transition displays strong interaction of configurations with varying orbital occupations, spin couplings, and crystal momentum of photoexcited electron across the first Brillouin zone.

\begin{figure*}[t!]
	\centering
	\captionsetup{justification=raggedright,singlelinecheck=false,font=footnotesize}
	\includegraphics[width=0.7\textwidth]{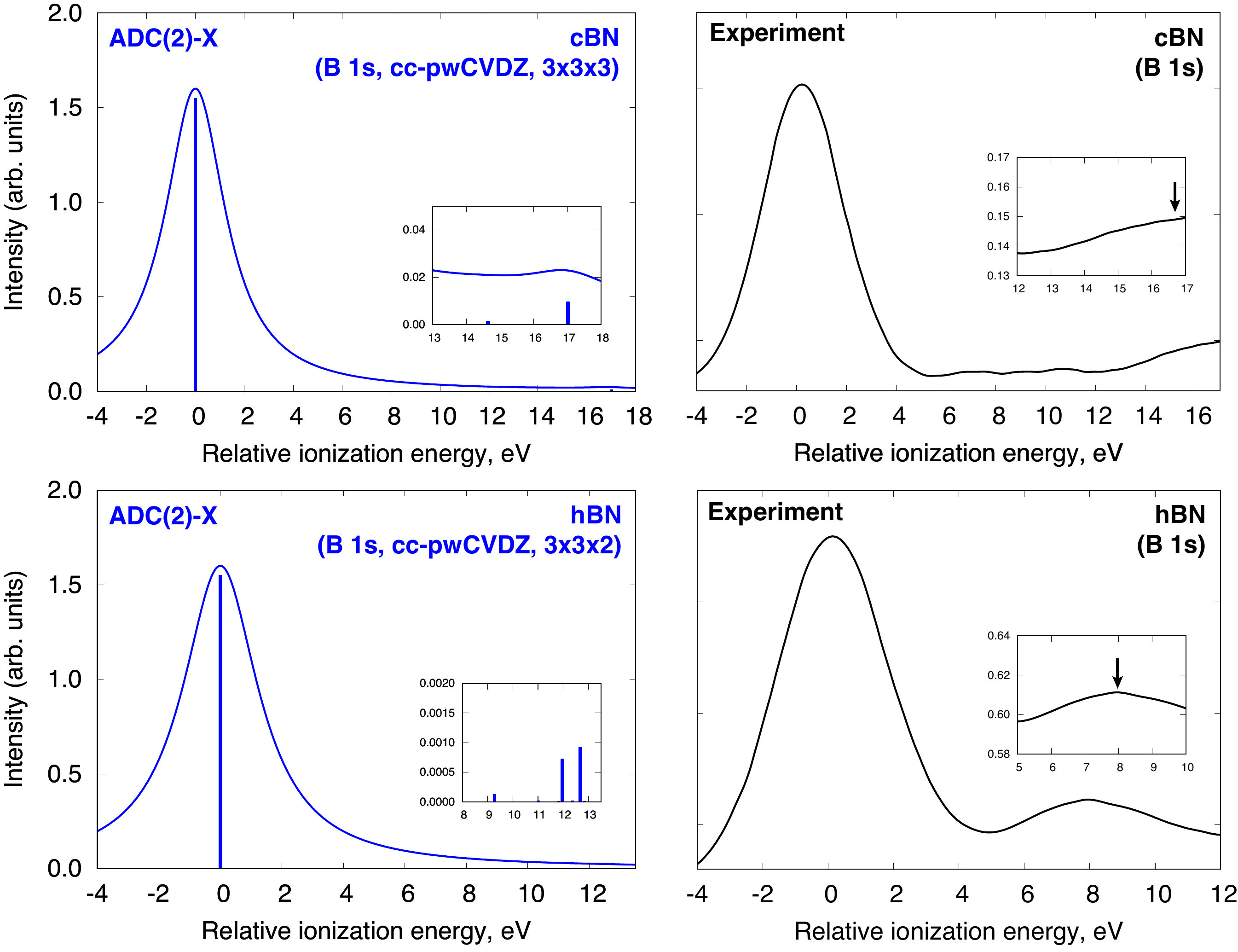}
	\caption{
		B 1s X-ray photoelectron spectra of cubic and hexagonal boron nitride (cBN and hBN) computed using ADC(2)-X with the cc-pwCVDZ basis set.
		For cBN and hBN, the 3$\times$3$\times$3 and 3$\times$3$\times$2 $k$-meshes were used, respectively.
		Also shown are experimental spectra from Ref.\@ \citenum{Deng.2006.10.1016/j.apsusc.2005.09.066} with satellite features indicated by arrows.
		To align the position of  first peak, the simulated and experimental spectra were shifted by $-$173.51 and $-$190.93 eV for cBN and by $-$184.15 and $-$190.77 eV for hBN, respectively.
		The experimental spectra from Ref.\@ \citenum{Deng.2006.10.1016/j.apsusc.2005.09.066} were reprinted with the permission of Elsevier.
	}
	\label{fig:bn_xps}
\end{figure*}

Next, we analyze the B 1s XPS spectra of cubic and hexagonal boron nitride (cBN and hBN) shown in \cref{fig:bn_xps}.
Although the B 1s ionization energies of two phases are very similar ($\sim$ 190.9 eV), the experimental XPS spectrum\cite{Deng.2006.10.1016/j.apsusc.2005.09.066} for hBN shows a distinct satellite feature $\sim$ 7.9 eV higher in energy than the B 1s edge.
The experimental cBN spectrum exhibits no shake-ups with significant intensity until $\sim$ 14 eV relative to the B 1s peak.\cite{Deng.2006.10.1016/j.apsusc.2005.09.066}
Consistent with experimental data, the ADC(2)-X calculations predict the lowest-energy satellites of cBN and hBN to be 14.64 and 9.27 eV, respectively.
Out of 140 transitions computed for cBN, only two shake-ups were found to have significant intensity with relative ionization energies of 14.64 and 17.02 eV.
These transitions correspond to excitations involving the three highest occupied and three lowest unoccupied band orbitals that are known to have the N 2p and B 2s/2p character, respectively. 
For hBN, several shake-ups with non-zero intensity were computed among the first 100 states.
The most prominent satellites appear at 9.27, 11.96, and 12.68 eV with 0.14, 0.78, and 1.0 relative intensities corresponding to $\pi\rightarrow\pi^*$ excitations between the two highest occupied and two lowest unoccupied band orbitals in a unit cell.
When compared to cBN, the satellites in hBN exhibit a stronger degree of configuration interaction, which is consistent with a more delocalized character of its band orbitals. 

\begin{figure*}[t!]
	\centering
	\captionsetup{justification=raggedright,singlelinecheck=false,font=footnotesize}
	\includegraphics[width=0.7\textwidth]{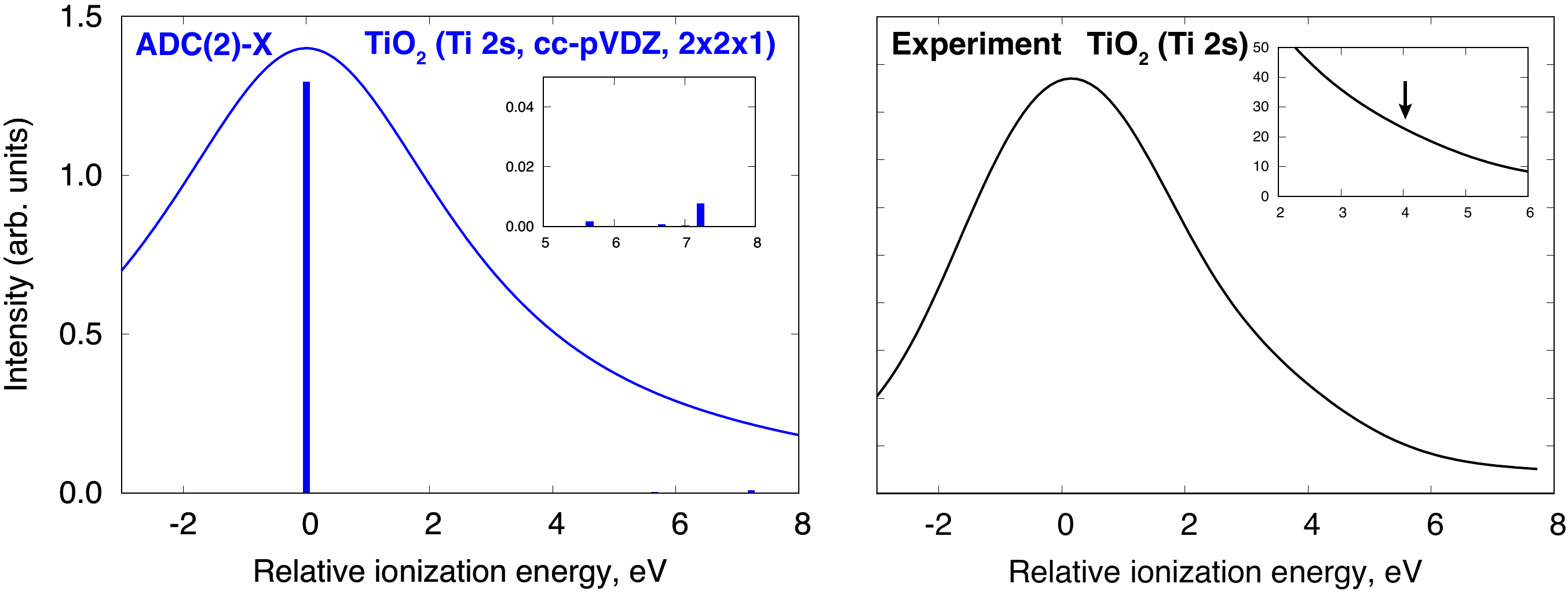}
	\caption{
		Ti 2s X-ray photoelectron spectrum of \ce{TiO2} computed using ADC(2)-X with the 2$\times$2$\times$1 $k$-mesh and the cc-pVDZ basis set.
		Also shown is experimental spectrum from Ref.\@ \citenum{Sen.1976.10.1016/0009-2614(76)80329-6} with satellite features indicated by arrows.
		To align the position of first peak, the simulated and experimental spectra were shifted by $-$553.12 and $-$565.52 eV, respectively.
		The experimental spectrum from Ref.\@ \citenum{Sen.1976.10.1016/0009-2614(76)80329-6} was reprinted with the permission of Elsevier.
	}
	\label{fig:tio2_xps}
\end{figure*}

Finally, we consider the Ti 2s XPS spectrum of rutile (\ce{TiO2}) that was reported to exhibit three satellites $\sim$ 4.0 $\pm$ 0.3, 13.5 $\pm$ 0.2, and 26.2 $\pm$ 0.3 eV above the Ti 2s edge in an experimental study by Sen and co-workers.\cite{Sen.1976.10.1016/0009-2614(76)80329-6}
Due to a high density of states, the ADC(2)-X calculation with 140 transitions provides spectral information for states with up to 7.8 eV in relative energy.
Thus, we limit our discussion to the first shake-up feature (4.0 $\pm$ 0.3 eV) that appears as a weak shoulder of the primary Ti 2s peak shown in \cref{fig:tio2_xps}.
The ADC(2)-X calculations performed using the cc-pVDZ basis set and 2$\times$2$\times$1 $k$-mesh predict several satellite transitions at 5.65, 6.04, 6.68, 7.01, and 7.22 eV with 0.21, 0.01, 0.08, 0.03, and 1.00 relative intensities, in a qualitative agreement with experimental data. 
Analysis of ADC(2)-X eigenvectors reveals that all shake-ups correspond to the ligand-to-metal charge transfer excitations with electrons being excited from the highest occupied band orbitals localized primarily on O atoms to the lowest unoccupied orbitals of predominantly Ti 3d character. 
In contrast to graphite and hBN, the orbital composition of satellite transitions in \ce{TiO2} is much more complex and delocalized, with at least 16 frontier band orbitals in a unit cell providing significant contributions to the ADC(2)-X eigenvectors. 

Overall, our results demonstrate that the ADC(2)-X method can simulate satellites in the XPS spectra of weakly correlated materials, providing useful insights into the nature of these elusive features. 
For all solids studied in this work, the relative energies of shake-ups are overestimated by $\sim$ 1 to 4 eV.
These findings are consistent with the molecular benchmarks of ADC(2)-X where it was shown to significantly overestimate the energies of doubly excited states due to approximate (first-order) treatment of electron correlation effects for double excitations.\cite{Dreuw:2014p82} 
Higher-order approximations in the ADC hierarchy are known to exhibit better results for doubly excited states, with ADC(4) expected achieve higher accuracy.\cite{Dobrodey:2002p022501,Thiel:2003p2088,Leitner.2024.10.1021/acs.jpca.4c03037}
Nevertheless, as a first-principles approach with a careful balance of computational cost and accuracy, ADC(2)-X can be used to improve the understanding of satellite transitions and many-body effects in the XPS spectra of materials.

%The computed satellite energies may also be affected by the finite-size and basis set incompleteness errors, although we believe these effects to be less significant when compared to the approximations in the description of electron correlation.
%For example, increasing the basis set from cc-pwCVDZ to cc-pVTZ lowers the energy of first satellite in the B 1s XPS spectrum of cBN by 0.15 eV.

\section{Conclusions}
\label{sec:conclusions}

Algebraic diagrammatic construction (ADC) is widely employed to simulate the interaction of molecules with high-energy (X-ray, XUV) light.
In this work, we present the first implementation and benchmark study of periodic ADC for calculating the core-ionized states and X-ray photoelectron spectra (XPS) of crystalline materials.
Our results demonstrate that the core ionization energies of elemental and binary solids (MgO, AlN, AlP, Si, BeO, cubic BN, SiC, diamond, GaN, and ZnO) computed using the extended second-order ADC approximation (ADC(2)-X) are within $\sim$ 0.5 eV of experimental measurements when using a triple-zeta Gaussian basis set, including valence band maximum and scalar relativistic corrections, and extrapolating to thermodynamic limit.
The strict second-order ADC method  (ADC(2)) was found to be significantly less accurate with a mean absolute error of $\sim$ 1.5 eV.
Crucially, the ADC(2)-X calculations do not rely on empirical or density functional parameters and can be used as benchmarks for lower-level theories when experimental data is not available.

Encouraged by our numerical results, we investigated if ADC(2)-X can simulate the satellite transitions in the XPS spectra of weakly correlated materials, namely: graphite, cubic and hexagonal boron nitrides, and \ce{TiO2}. 
For all considered solids, ADC(2)-X was able to predict satellites with noticeable intensities but significantly (by $\sim$ 1 to 4 eV) overestimated their relative energies.
Analyzing the eigenvectors of ADC(2)-X effective Hamiltonian revealed that the satellites exhibit strong interaction of electronic configurations across the first Brillouin zone, providing insight into the many-body interactions that underpin these elusive XPS features.

Our work motivates several important research directions.
To enable the ADC(2)-X calculations for more complex materials with larger basis sets and smaller finite-size errors, the efficiency of its computer implementation needs to be further improved.
This can be achieved by taking advantage of local correlation techniques, tensor decomposition, and developing accurate approaches for extrapolating the computed energies to thermodynamic limit. 
Our current research also motivates further developments in periodic many-body theory to enable the accurate first-principles calculations of satellite states in solids with a wide range of electron correlations.
Here, the ADC framework may prove to be useful in analyzing the importance of diagrammatic contributions and developing approximations that can accurately describe satellites with affordable computational cost.
Finally, the periodic ADC methods can be further extended to simulate a broader range of spectroscopic observables, enhancing the understanding of electron correlation effects in materials at high excitation energies by bridging the gap between theory and experiments.
Expanding these horizons will unlock new opportunities to deepen our understanding of electronic structure in condensed phases.

\suppinfo
Structural parameters, valence band maximum ionization energies, scalar relativistic corrections, benchmarking of the mixed (cc-pwCVTZ + cc-pVDZ) basis set, and the results of ADC(2)-X calculations for the satellite states in XPS spectra.

\acknowledgement
This research was supported by the Exploration Grant Program of the College of Arts and Sciences at the Ohio State University and the donors of the American Chemical Society Petroleum Research Fund (PRF grant no.\@ 65903-ND6).
Computations were performed at the Ohio Supercomputer Center under the project PAS1963.\cite{OhioSupercomputerCenter1987}

%\bibliography{refs_alex,papers_alex}

\providecommand{\latin}[1]{#1}
\makeatletter
\providecommand{\doi}
  {\begingroup\let\do\@makeother\dospecials
  \catcode`\{=1 \catcode`\}=2 \doi@aux}
\providecommand{\doi@aux}[1]{\endgroup\texttt{#1}}
\makeatother
\providecommand*\mcitethebibliography{\thebibliography}
\csname @ifundefined\endcsname{endmcitethebibliography}
  {\let\endmcitethebibliography\endthebibliography}{}

\end{document}